\documentclass[twocolumn]{emulateapj}
\usepackage{apjfonts}
\usepackage{natbib}
\newcommand{\etcn}{C$_2$H$_5$CN}
\newcommand{\be}{\begin{equation}}
\newcommand{\ee}{\end{equation}}

\newcommand{\eg}{\emph{e.g.}}
\newcommand{\kms}{\mbox{km\ \ensuremath{\rm{s}^{-1}}}}

\shortauthors{Cordiner et al.}

\begin{document}

\title{Ethyl cyanide on Titan: Spectroscopic detection and mapping using ALMA}

\submitted{Published in ApJL, 800, L14}

\author{M. A. Cordiner\altaffilmark{1,2}, M. Y. Palmer\altaffilmark{1,3}, C. A. Nixon\altaffilmark{1}, P. G. J. Irwin\altaffilmark{4}, N. A. Teanby\altaffilmark{5}, S. B. Charnley\altaffilmark{1}, M. J. Mumma\altaffilmark{1}, Z. Kisiel\altaffilmark{6}, J. Serigano\altaffilmark{1,2}, Y.-J. Kuan\altaffilmark{7,8}, Y.-L. Chuang\altaffilmark{7}, K.-S. Wang\altaffilmark{8}}

\altaffiltext{1}{NASA Goddard Space Flight Center, 8800 Greenbelt Road, Greenbelt, MD 20771, USA.}
\email{martin.cordiner@nasa.gov}
\altaffiltext{2}{Department of Physics, Catholic University of America, Washington, DC 20064, USA.}
\altaffiltext{3}{Department of Chemistry, St. Olaf College, 1520 St. Olaf Avenue, Northfield, MN 55057, USA.}
\altaffiltext{4}{Atmospheric, Oceanic and Planetary Physics, Clarendon Laboratory, University of Oxford, Parks Road, Oxford, OX1 3PU, UK.}
\altaffiltext{5}{School of Earth Sciences, University of Bristol, Wills Memorial Building, Queen's Road, Bristol, BS8 1RJ, UK.}
\altaffiltext{6}{Institute of Physics, Polish Academy of Sciences, Al. Lotnik{\o}w 32/46, 02-668 Warszawa, Poland.}
\altaffiltext{7}{National Taiwan Normal University, Taipei 116, Taiwan, ROC.}
\altaffiltext{8}{Institute of Astronomy and Astrophysics, Academia Sinica, Taipei 106, Taiwan, ROC.}

\begin{abstract}

We report the first spectroscopic detection of ethyl cyanide (\etcn) in Titan's atmosphere, obtained using spectrally and spatially resolved observations of multiple emission lines with the Atacama Large Millimeter/submillimeter Array (ALMA).  The presence of \etcn\ in Titan's ionosphere was previously inferred from Cassini ion mass spectrometry measurements of C$_2$H$_5$CNH$^+$. Here we report the detection of 27 rotational lines from \etcn\ (in 19 separate emission features detected at $>3\sigma$ confidence), in the frequency range 222-241~GHz. Simultaneous detections of multiple emission lines from HC$_3$N, CH$_3$CN  and CH$_3$CCH were also obtained. In contrast to HC$_3$N, CH$_3$CN and CH$_3$CCH, which peak in Titan's northern (spring) hemisphere, the emission from \etcn\ is found to be concentrated in the southern (autumn) hemisphere, suggesting a distinctly different chemistry for this species, consistent with a relatively short chemical lifetime for \etcn. Radiative transfer models show that \etcn\ is most concentrated at altitudes $\gtrsim200$~km, suggesting production predominantly in the stratosphere and above. Vertical column densities are found to be in the range (1-5)$\times10^{14}$~cm$^{-2}$. 

\end{abstract}

\keywords{planets and satellites: individual (Titan) --- planets and satellites: atmospheres --- techniques: interferometric --- techniques: imaging spectroscopy}

\section{Introduction}

Saturn's largest moon Titan has a thick (1.45~bar) atmosphere composed primarily of molecular nitrogen ($98$\%) and methane (approximately $2$\%). Remote and in-situ measurements \citep[see the review by][]{bez14}, have {also} revealed the presence of a diverse population of trace hydrocarbons and nitrogen-bearing compounds, the presence of which can be explained as a consequence of a complex gas-phase photochemistry, driven by photodissociation/ionization and cosmic rays \citep{wil04}. 

Multiple studies have predicted the existence of ethyl cyanide (\etcn, also known as propionitrile) on Titan, but a firm confirmation of its presence has been elusive until now. \citet{cap81} suggested that \etcn\ could be formed as a recombination product of complex nitrile ions produced in three-body association reactions in the dense, lower atmosphere. More recent photochemical models by \citet{kra09} and \citet{loi14} predict \etcn\ to be abundant throughout the atmosphere. In laboratory plasma-discharge experiments designed to simulate Titan's photochemistry, \etcn\ is among the most abundant molecules produced \citep[\eg][]{tho91,col99,fuj99}, and crystalline C$_2$H$_5$CN has been suggested as an explanation for the broad emission feature at 221~cm$^{-1}$ seen in the Voyager IRIS data \citep{kha05}. Nevertheless, previous spectroscopic studies have failed to {find} gas-phase ethyl cyanide; abundance upper limits have been obtained of $8\times10^{-9}$ from the Cassini Composite Infrared Spectrometer (CIRS) and $2\times10^{-9}$ from IRAM 30-m microwave spectroscopy (\citealt{dek08} and \citealt{mar02}, respectively).

The presence of \etcn\ was strongly implied by the detection of C$_2$H$_5$CNH$^+$ by the Cassini Ion and Neutral Mass Spectrometer (INMS) \citep{vui07}. This nitrile ion was theorized to form primarily by proton transfer from HCNH$^+$ and C$_2$H$_5$$^+$ to neutral \etcn. A spectroscopic detection of \etcn\ would provide the first conclusive proof for the presence of ethyl cyanide in Titan's atmosphere and help validate the conclusions of the INMS studies. Gas-phase synthesis of C$_2$H$_5$CN is not well understood, and the chemical models of \citet{kra09} and \citet{loi14} differ in their predicted abundances by 2-3 orders of magnitude. Accurate measurements of the \etcn\ atmospheric distribution are required in order to constrain models for the formation of nitriles and other large organic molecules, which will lead to improvements in our understanding of the complex photochemistry occurring on Titan and on other bodies with nitrogen and methane-rich atmospheres.

In this study, we employ millimeter-wave data obtained from the Atacama Large Millimeter/submillimeter Array (ALMA) Science Archive\footnote{https://almascience.nrao.edu/alma-data/archive} to search for and map rotational emission lines from C$_2$H$_5$CN in Titan's atmosphere, and compare the spatial distribution of \etcn\ emission with that of other, previously-identified molecules.

\section{Observations}
\label{obs}

Interferometric observations of Titan were made between UT 2012-07-03 23:22:14 and 2012-07-04 01:06:18 as part of ALMA project 2011.0.00319.S {(for the purpose of flux calibration).}  Two 4 minute integrations were obtained using the dual-sideband Band 6 receiver, with twenty 12-m antennae in the telescope array, providing baselines in the range 21-402~m. {Due to the high sensitivity of ALMA, even these short integrations permitted the detection of weak spectral lines on Titan.} The correlator was configured to observe four basebands, with frequencies 221.48-223.35~GHz and 223.48-225.35~GHz in the lower sideband, and 236.48-238.36~GHz and 239.48-241.36~GHz in the upper sideband, with 3840 channels per spectral window. The channel spacing was 488~kHz which (after Hanning smoothing by the correlator), leads to a spectral resolution of 976~kHz (or $1.3$~\kms\ at 224~GHz).  Weather conditions were good, with 1.7~mm of precipitable water vapor at zenith. The bright quasar 3C\,279 was observed for bandpass calibration.  The telescope was configured to track Titan's ephemeris position, updating the coordinates of the phase center in real-time.

The data obtained from the ALMA Science Archive were already processed using the standard scripts and protocols provided by the Joint ALMA Observatory. {This included routine flagging, bandpass calibration and complex gain calibration.} {The measured continuum flux density for each baseline} was scaled to match the Butler-JPL-Horizons 2010 Titan flux model (see ALMA Memo \#594\footnote{https://science.nrao.edu/facilities/alma/aboutALMA/Technology/ALMA\\\_Memo\_Series/alma594/memo594.pdf}), which is expected to be accurate to within 15\%.  Continuum-subtraction of the visibility amplitudes was performed using the {\tt uvcontsub} task in the NRAO CASA software (version 4.2.1) \citep{mcm07}, and imaging was carried out using the {\tt clean} task.   Deconvolution of the point-spread function (PSF) was performed for each spectral channel using the Hogbom algorithm, with natural visibility weighting and a threshold flux level of twice the {expected} RMS noise. The image pixel sizes were set to $0.1''\times0.1''$. The resulting spatial resolution (FWHM of the Gaussian restoring beam at 223.5~GHz) was $0.85''\times0.67''$ (long axis $57.3^{\circ}$ clockwise from celestial north). This resolution element corresponds to $5860\times4620$~km at Titan's geocentric distance of 9.51~AU at the time of observation (compared with Titan's 5,150~km diameter). 

The images were transformed from equatorial coordinates to (projected) {linear} distances with respect to the center of Titan, and the spectral coordinate scale {of each image} was Doppler-shifted to Titan's rest frame using the JPL Horizons {Topocentric radial velocity}.

\begin{figure}
\centering
\includegraphics[width=0.85\columnwidth]{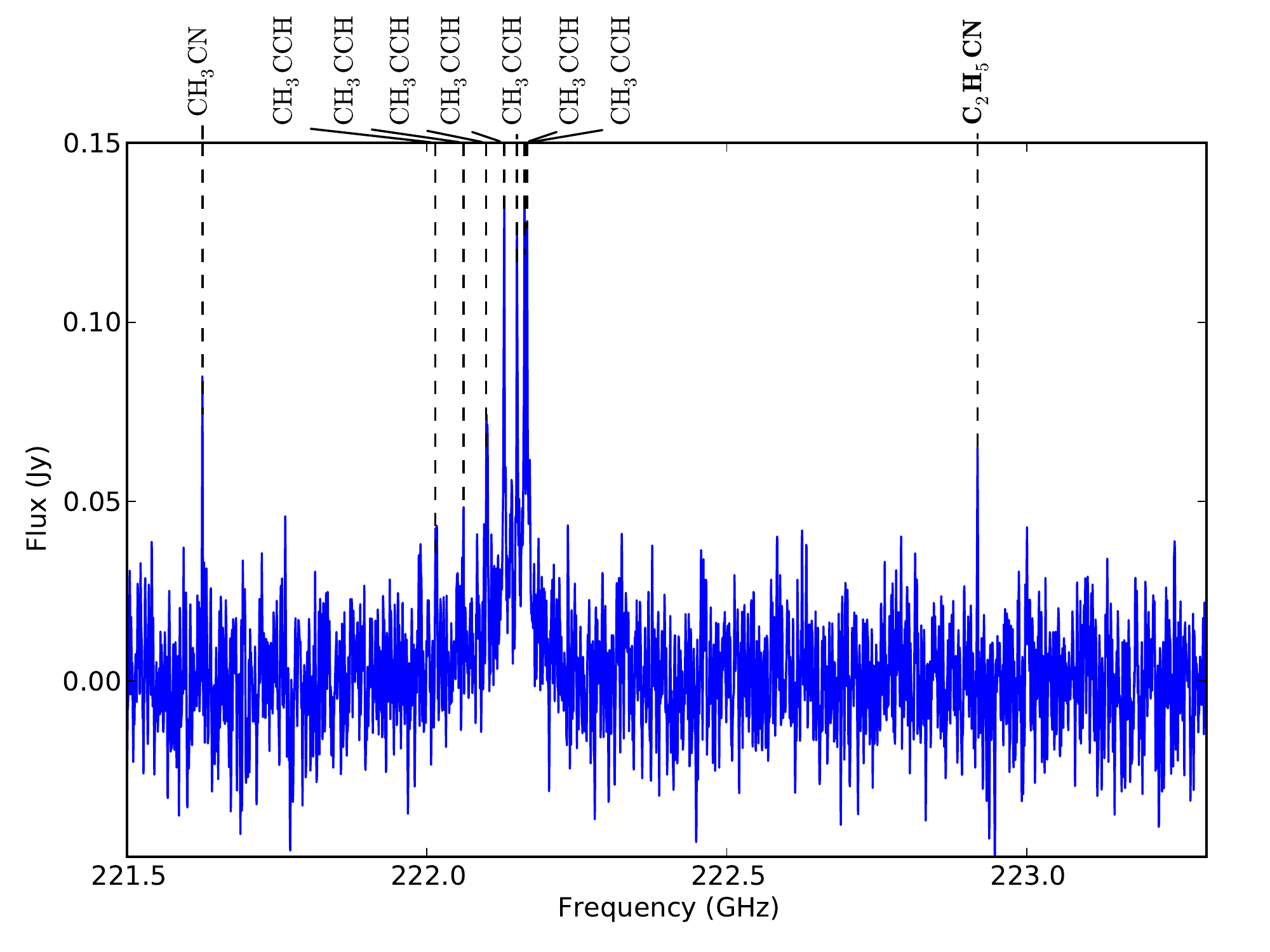}
\includegraphics[width=0.85\columnwidth]{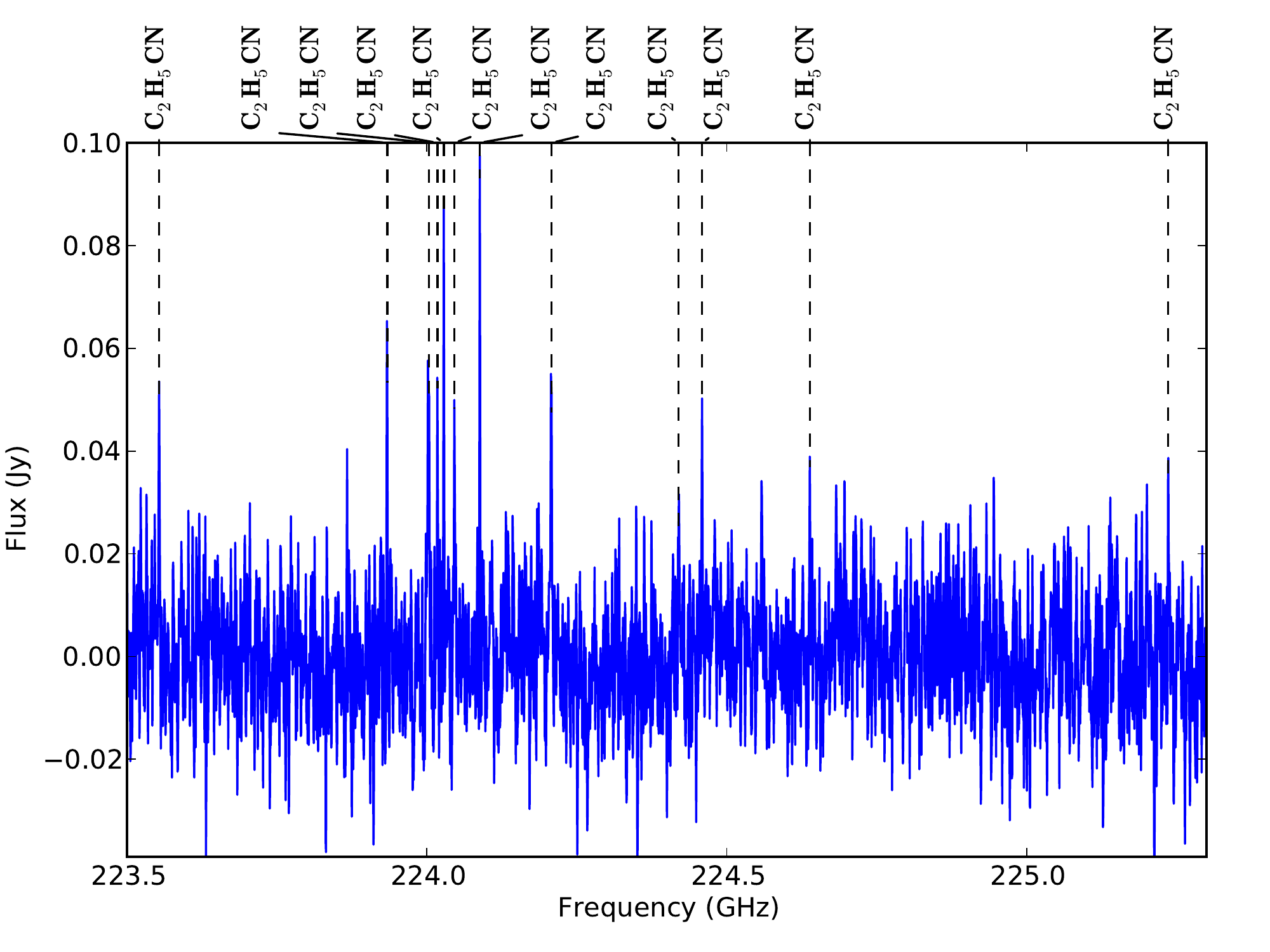}\\
\includegraphics[width=0.85\columnwidth]{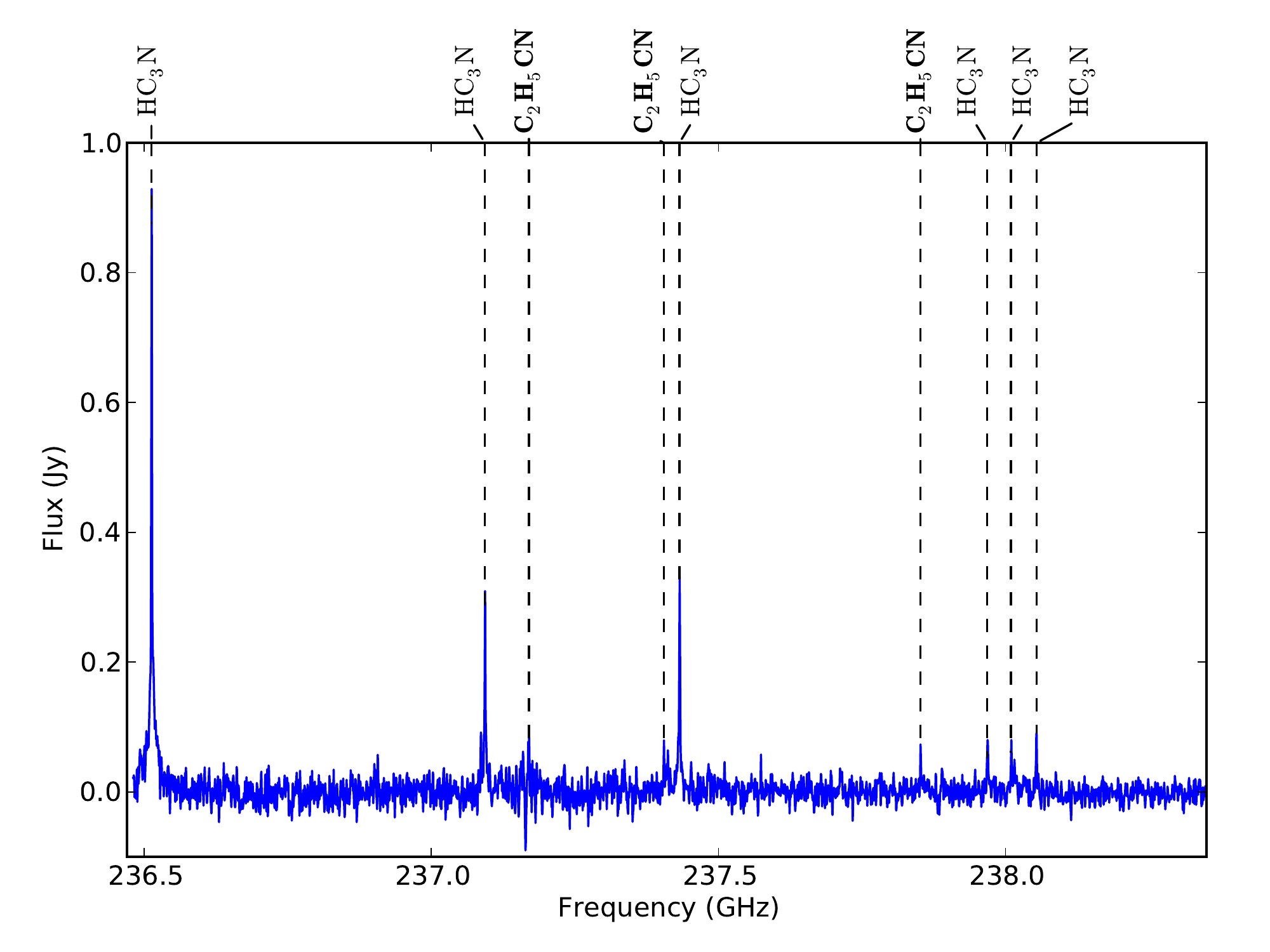}
\includegraphics[width=0.85\columnwidth]{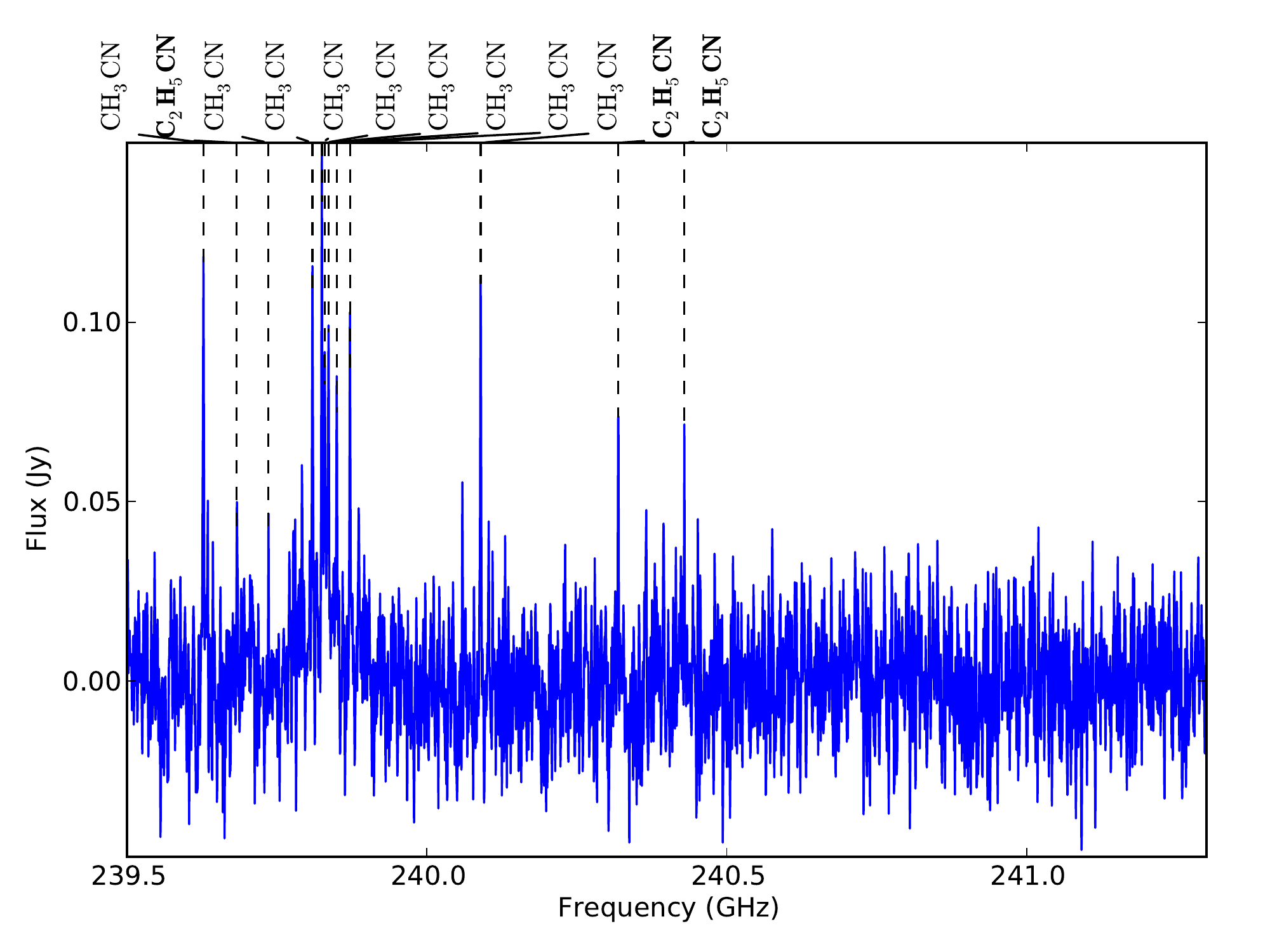}
\caption{ALMA spectra observed 2012 July 3-4, integrated over a circular {region of radius $1.1''$ centered on Titan}. Titan's continuum emission has been subtracted. Prominent emission lines are assigned by species; bold typeface highlights the \etcn\ emission features. The small absorption feature at 237.146~GHz is attributed to ozone in the Earth's atmosphere. \label{fig:spectra} }
\end{figure}

\begin{table}
\centering
\caption{Detected transitions and upper-state energies\label{tab:trans}}
{\footnotesize
\begin{tabular}{lcr@{$-$}lr}
\hline\hline
Species&Frequency (GHz)&\multicolumn{2}{c}{Transition}&$E_u$ (K)\\
\hline
C$_2$H$_5$CN & 222.918      & $25_{1,24}       $&$ 24_{1,23}      $ &     142\\
C$_2$H$_5$CN & 223.554      & $26_{0,26}       $&$ 25_{0,25}      $ &     147\\
C$_2$H$_5$CN & 223.934      & $25_{3,23}       $&$ 24_{3,22}      $ &     150\\
C$_2$H$_5$CN & 224.002      & $25_{10,15}       $&$ 24_{10,14}      $ &     251\\
C$_2$H$_5$CN & 224.002      & $25_{10,16}       $&$ 24_{10,15}      $ &     251\\
C$_2$H$_5$CN & 224.003      & $25_{9,17}       $&$ 24_{9,16}      $ &     230\\
C$_2$H$_5$CN & 224.003      & $25_{9,16}       $&$ 24_{9,15}      $ &     230\\
C$_2$H$_5$CN & 224.018      & $25_{11,14}       $&$ 24_{11,13}      $ &     274\\
C$_2$H$_5$CN & 224.018      & $25_{11,15}       $&$ 24_{11,14}      $ &     274\\
C$_2$H$_5$CN & 224.028      & $25_{8,18}       $&$ 24_{8,17}      $ &     211\\
C$_2$H$_5$CN & 224.028      & $25_{8,17}       $&$ 24_{8,16}      $ &     211\\
C$_2$H$_5$CN & 224.046      & $25_{12,13}       $&$ 24_{12,12}      $ &     300\\
C$_2$H$_5$CN & 224.046      & $25_{12,14}       $&$ 24_{12,13}      $ &     300\\
C$_2$H$_5$CN & 224.088      & $25_{7,19}       $&$ 24_{7,18}      $ &     194\\
C$_2$H$_5$CN & 224.088      & $25_{7,18}       $&$ 24_{7,17}      $ &     194\\
C$_2$H$_5$CN & 224.207      & $25_{6,20}       $&$ 24_{6,19}      $ &     180\\
C$_2$H$_5$CN & 224.208      & $25_{6,19}       $&$ 24_{6,18}      $ &     180\\
C$_2$H$_5$CN & 224.420      & $25_{5,21}       $&$ 24_{5,20}      $ &     168\\
C$_2$H$_5$CN & 224.459      & $25_{5,20}       $&$ 24_{5,19}      $ &     168\\
C$_2$H$_5$CN & 224.639      & $25_{4,22}       $&$ 24_{4,21}      $ &     158\\
C$_2$H$_5$CN & 225.236      & $25_{4,21}       $&$ 24_{4,20}      $ &     158\\
C$_2$H$_5$CN & 237.170      & $26_{3,23}       $&$ 25_{3,22}      $ &     162\\
C$_2$H$_5$CN & 237.405      & $26_{2,24}       $&$ 25_{2,23}      $ &     158\\
C$_2$H$_5$CN & 237.852      & $27_{2,26}       $&$ 26_{2,25}      $ &     165\\
C$_2$H$_5$CN & 239.683      & $27_{1,26}       $&$ 26_{1,25}      $ &     165\\
C$_2$H$_5$CN & 240.319      & $28_{1,28}       $&$ 27_{1,27}      $ &     169\\
C$_2$H$_5$CN & 240.429      & $28_{0,28}       $&$ 27_{0,27}      $ &     169\\[2mm]
HC$_3$N   & 236.513      & $26           $&$ 25          $ &     153\\
HC$_3$N $v_7=1$  & 237.093      & $26           $&$ 25\ 1e          $ &     474\\
HC$_3$N $v_7=1$  & 237.432      & $26           $&$ 25\ 1f          $ &     474\\
HC$_3$N $v_7=2$  & 237.968      & $26           $&$ 25\ 0e          $ &     792\\
HC$_3$N $v_7=2$  & 238.010      & $26           $&$ 25\ 2e         $ &     795\\
HC$_3$N $v_7=2$  & 238.054      & $26           $&$ 25\ 2f         $ &     795\\[2mm]
CH$_3$CN $v_8=1$ & 221.626   & $12_1$&$11_{-1}$                   &     588\\
CH$_3$CN $v_8=1$ & 239.627   & $13_1$&$12_{-1}$                   &     600\\
CH$_3$CN $v_8=1$ & 239.736   & $13_4$&$12_{4}$                   &      773\\
CH$_3$CN $v_8=1$ & 239.808   & $13_2$&$12_{2}$                   &      661\\
CH$_3$CN $v_8=1$ & 239.825   & $13_4$&$12_{-4}$                   &      667\\
CH$_3$CN $v_8=1$ & 239.830   & $13_1$&$12_{1}$                   &      626\\
CH$_3$CN $v_8=1$ & 239.836   & $13_0$&$12_{0}$                   &      606\\
CH$_3$CN $v_8=1$ & 239.850   & $13_3$&$12_{-3}$                   &      630\\
CH$_3$CN $v_8=1$ & 239.872   & $13_2$&$12_{-2}$                   &      608\\
CH$_3$CN $v_8=1$ & 240.090   & $13_{-1}$&$12_{1}$               &      600\\[2mm]
CH$_3$CCH & 222.014      & $13_6         $&$ 12_6        $ &     334\\
CH$_3$CCH & 222.061      & $13_5         $&$ 12_5        $ &     255\\
CH$_3$CCH & 222.099      & $13_4         $&$ 12_4        $ &     190\\
CH$_3$CCH & 222.129      & $13_3         $&$ 12_3        $ &     139\\
CH$_3$CCH & 222.150      & $13_2         $&$ 12_2        $ &     103\\
CH$_3$CCH & 222.163      & $13_1         $&$ 12_1        $ &      82\\
CH$_3$CCH & 222.167      & $13_0         $&$ 12_0        $ &      75\\
\hline
\end{tabular}
}
\parbox{0.9\columnwidth}{\footnotesize 
\vspace*{1mm}
{\bf Note.} Primary spectroscopic sources for molecular line frequencies: C$_2$H$_5$CN --- \citet{bra09}, HC$_3$N --- \citet{tho00}, CH$_3$CN --- \citet{bou80}, CH$_3$CCH --- \citet{dub78}.}

\end{table}

\begin{figure*}
\centering
\includegraphics[width=0.4\textwidth]{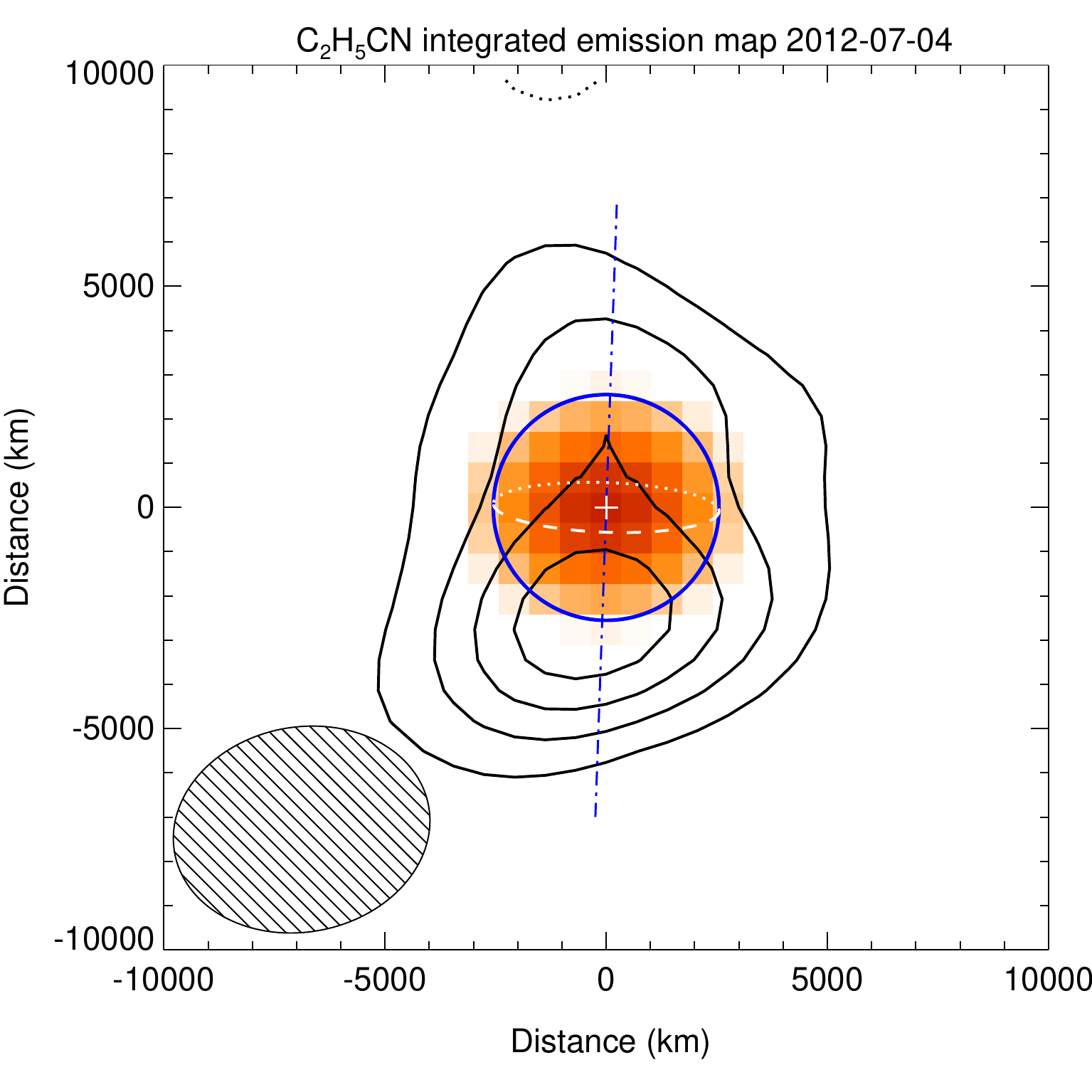}
\hspace{10mm}
\includegraphics[width=0.4\textwidth]{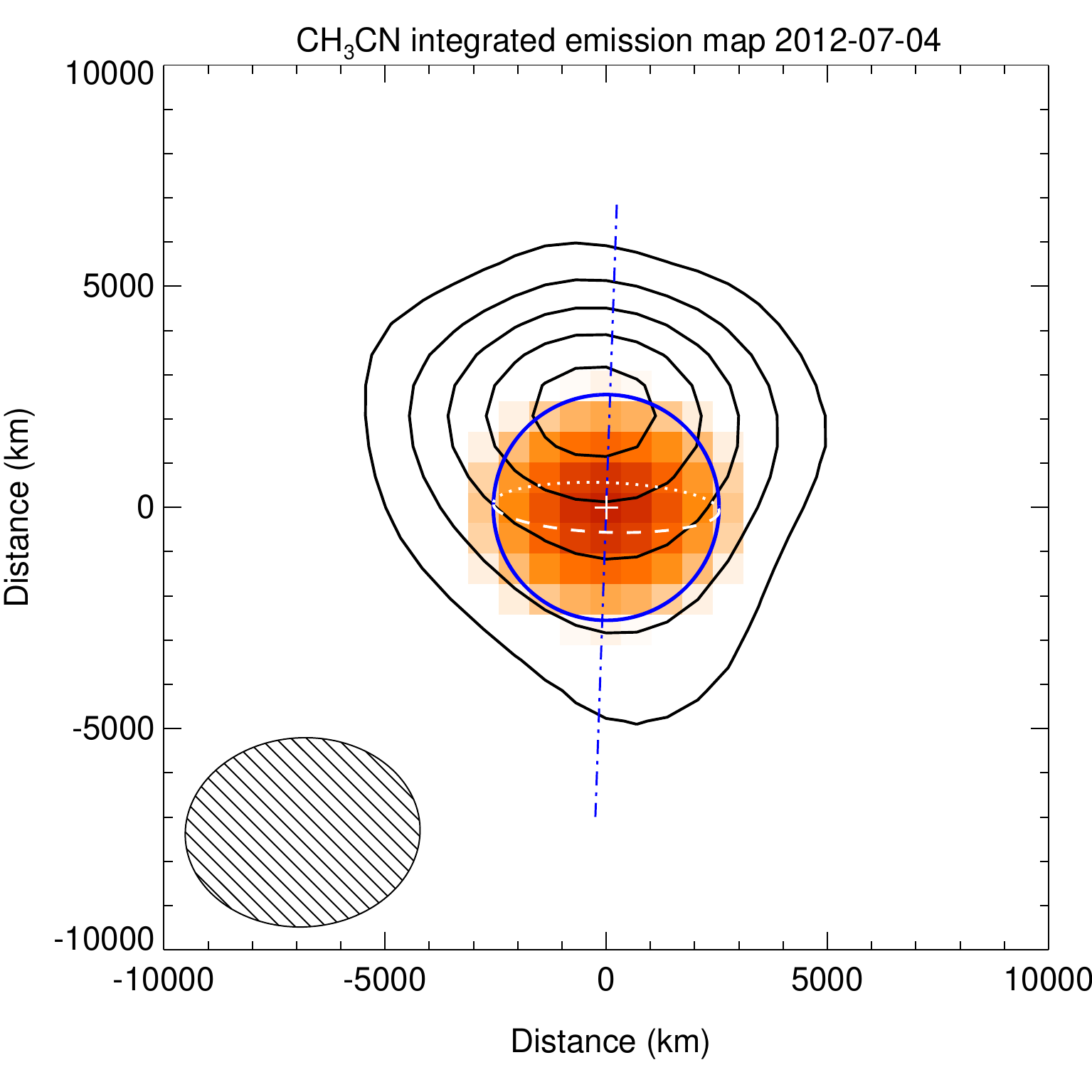}\\
\includegraphics[width=0.4\textwidth]{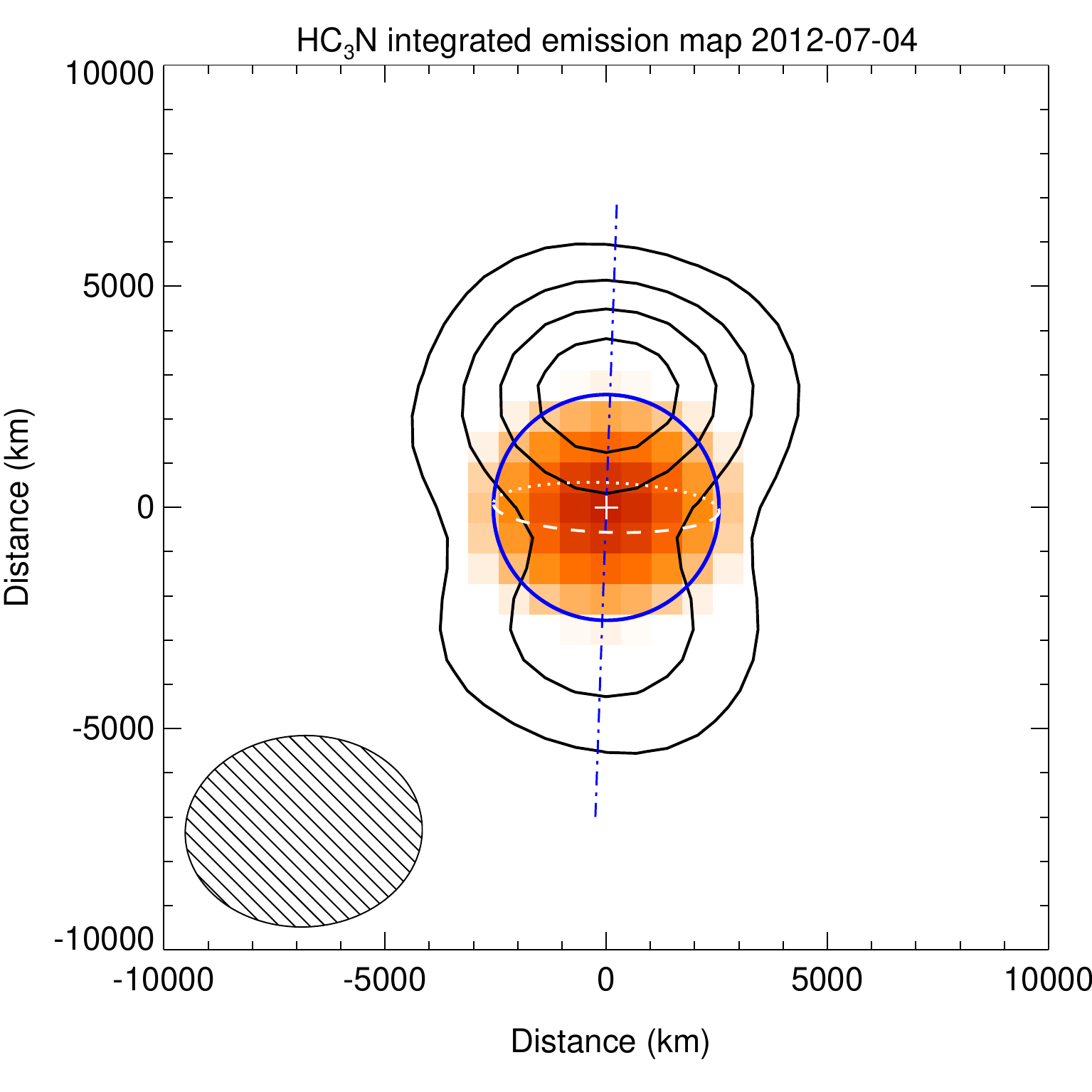}
\hspace{10mm}
\includegraphics[width=0.4\textwidth]{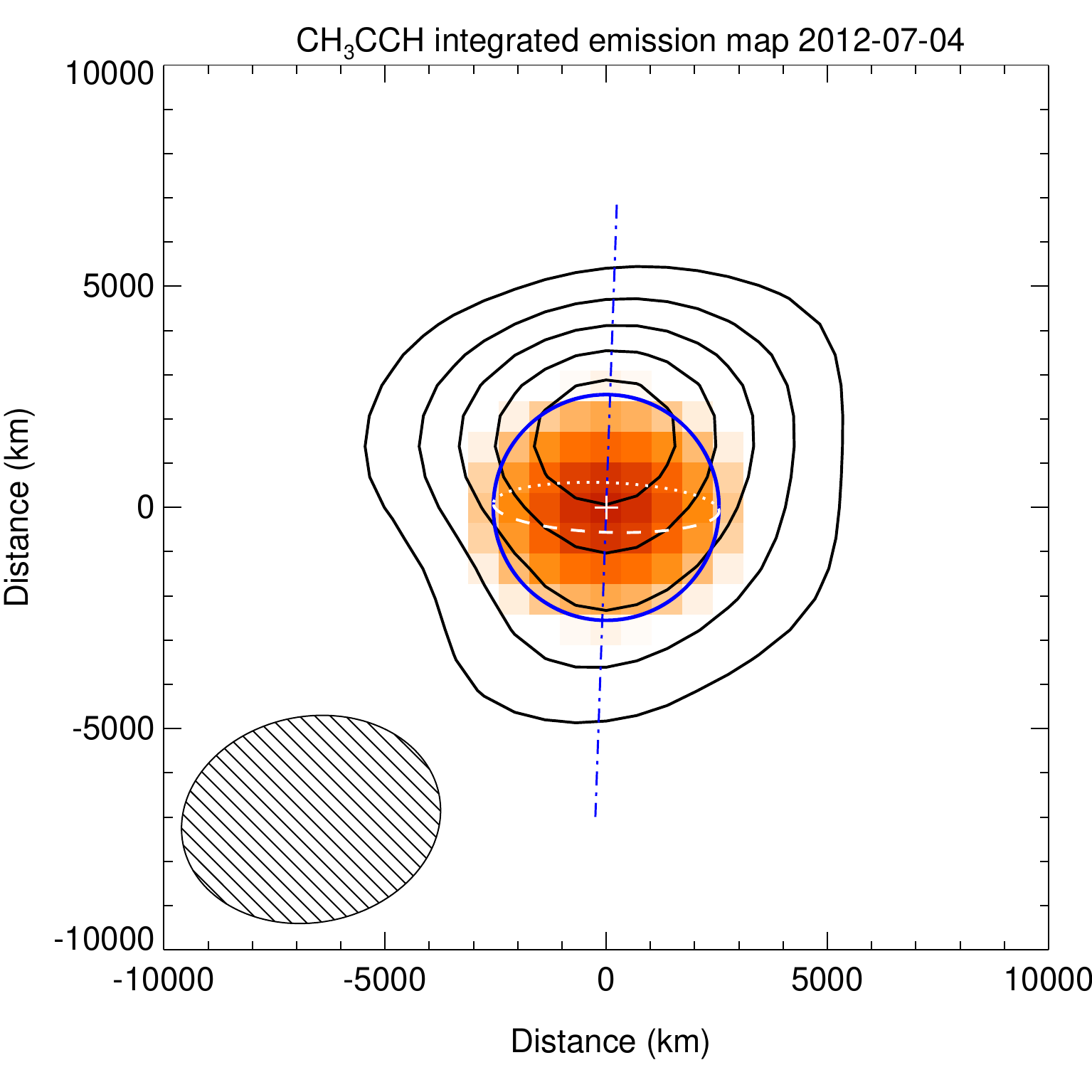}
\caption{Integrated emission contour maps for \etcn, HC$_3$N, CH$_3$CN and CH$_3$CCH. The 221-225~GHz continuum is shown in orange. The coordinate scale is in Titan-projected distances (white cross denotes the {position of the phase center}). Axes are aligned in the equatorial coordinate system. {Contour intervals (in units of $\sigma$ --- the RMS noise level of each map), are as follows: \etcn: $3\sigma$, HC$_3$N: $8\sigma$, CH$_3$CN: $3\sigma$, CH$_3$CCH: $5\sigma$.} The blue circle represents Titan's surface (dashed white curve is the equator), and the dot-dashed blue line is the polar axis, oriented 2.0$^{\circ}$ clockwise from vertical, with the north pole tilted toward the observer by $12.7^{\circ}$. Titan's Earth-facing hemisphere was almost fully illuminated at the time of observation, with a (Sun-target-observer) phase angle of $5.9^{\circ}$. {The
FWHM of the Gaussian restoring beam ($0.85''\times0.67''$) and its orientation are shown as hatched ellipses}.\label{fig:maps}}
\end{figure*}

\section{Results}
\label{results}

The observed spectra recorded in each of the four basebands are shown in Fig. \ref{fig:spectra}. These were {obtained} by integrating the reduced ALMA data cubes {within a circular region of radius $1.1''$ from the center of Titan} (7,500~km at Titan's distance). This {circle} enclosed the flux within $2\sigma_{PSF}$ of Titan's surface, where $\sigma_{PSF}$ is the standard deviation of the major axis of the {restoring beam}. Negligible flux was found to lie outside of this {region}.

Spectral peaks were assigned using the Splatalogue database for astronomical spectroscopy\footnote{http://www.cv.nrao.edu/php/splat/}, with frequencies {obtained} from the Cologne Database for Molecular Spectroscopy \citep{mul01} and the JPL catalog\footnote{http://spec.jpl.nasa.gov/}. The detected transitions are given in Table \ref{tab:trans}. The spectra are sufficiently uncrowded that unambiguous assignments were possible for all detected lines. Nineteen separate \etcn\ rotational emission features (many of which are blends of several transitions) were detected at greater than $3\sigma$ confidence. Emission from HC$_3$N, vibrationally-excited CH$_3$CN and CH$_3$CCH was also detected. An additional 153 \etcn\ lines lie within the observed spectral range, but these were too weak to be detected {individually}.

\begin{figure}
\vspace*{-0.5cm}
\centering
\includegraphics[width=\columnwidth]{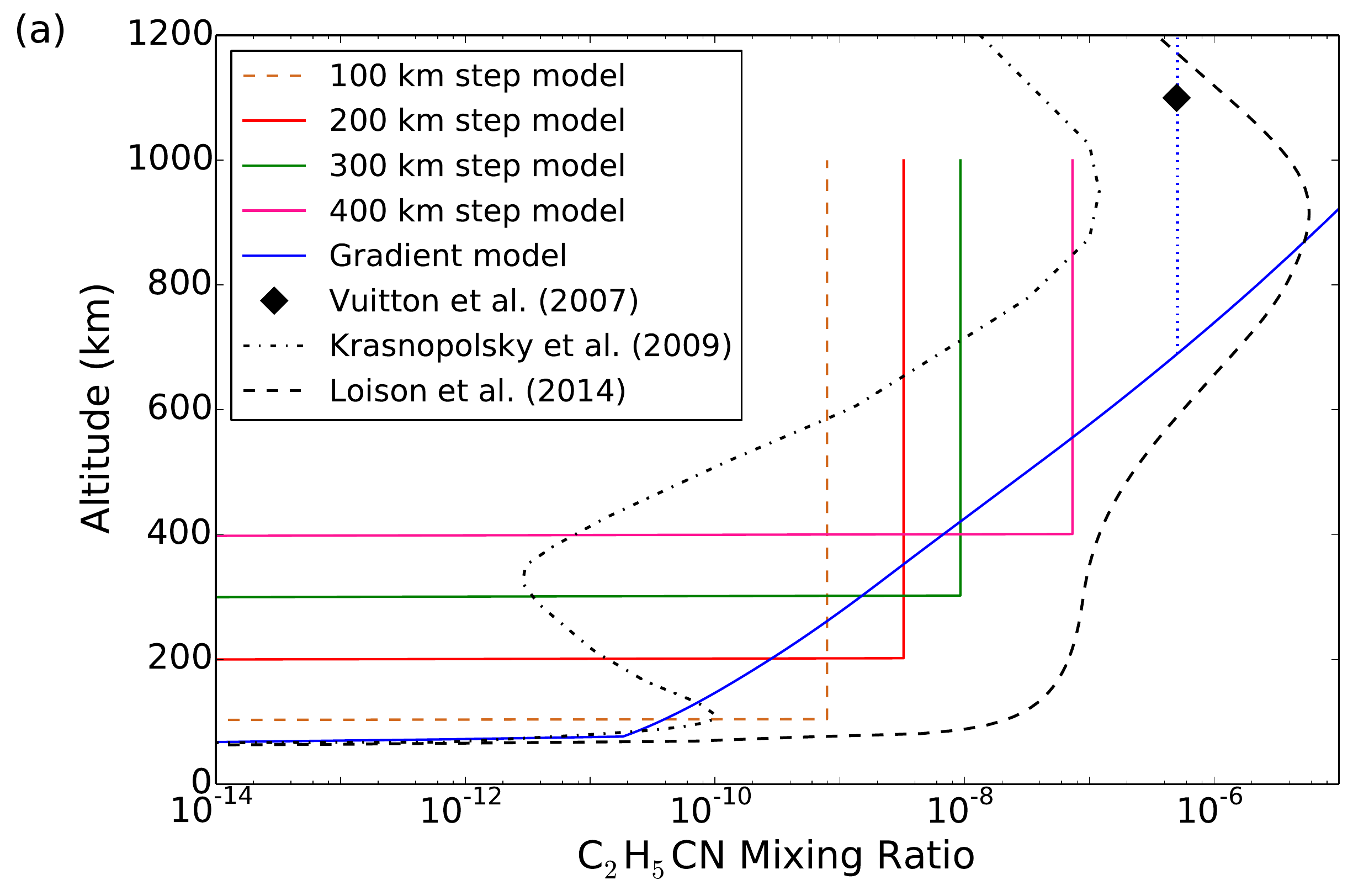}
\includegraphics[width=\columnwidth]{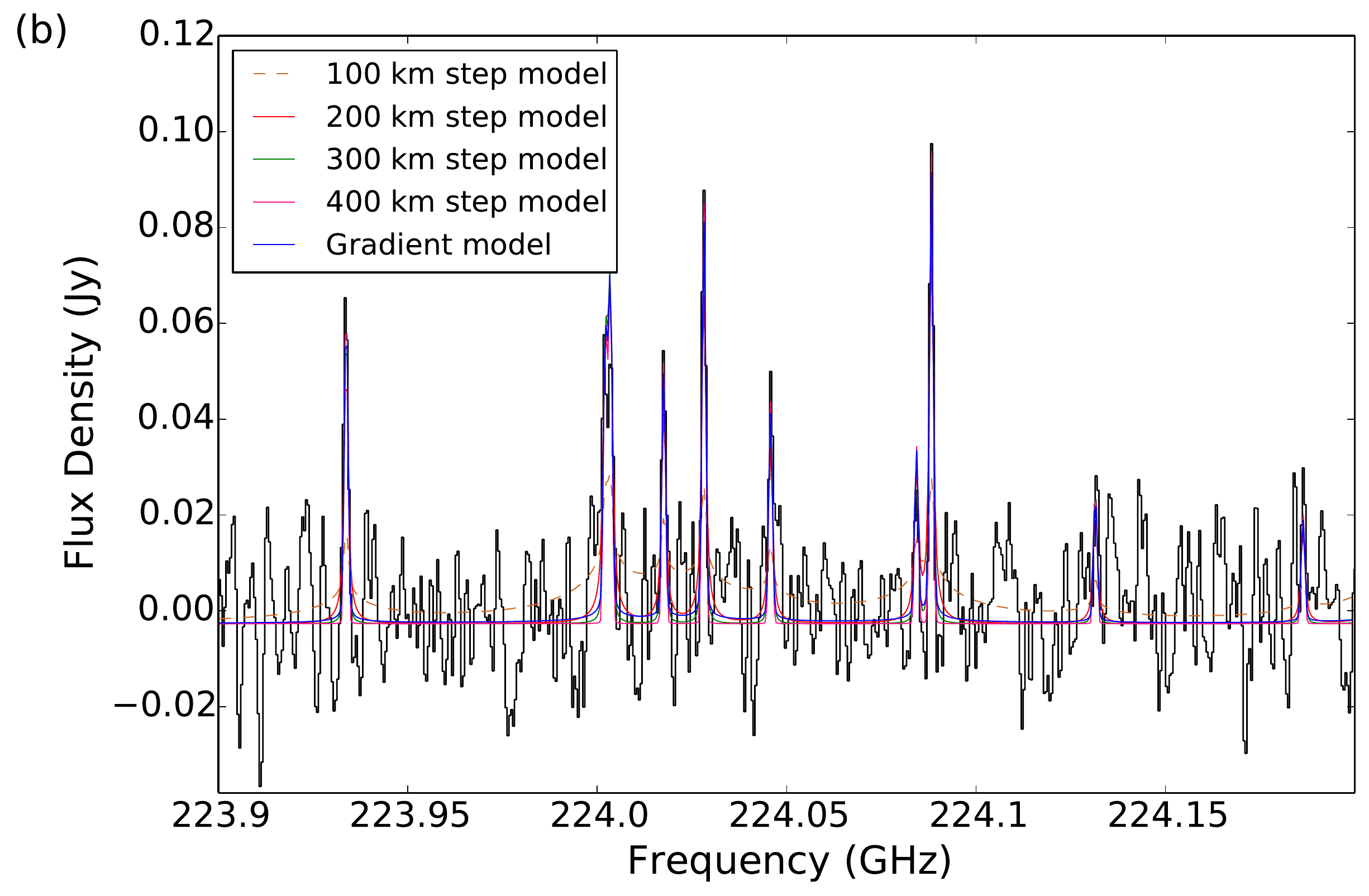}
\includegraphics[width=\columnwidth]{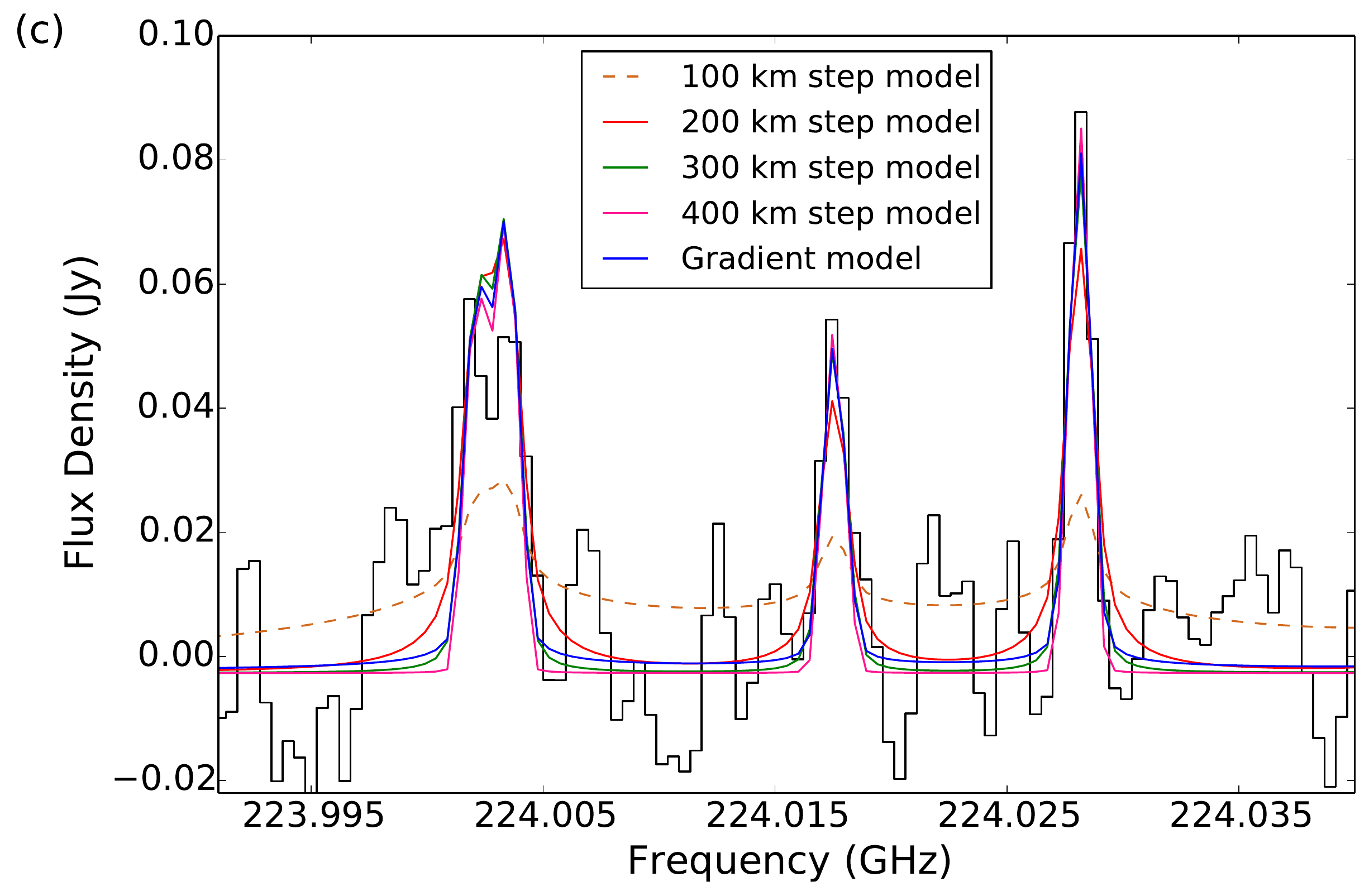}
\caption{(a) VMR profiles for the best-fitting \etcn\ models (parameters given in Table \ref{tab:models}). {The (poorly-fitting) 100~km step model is shown}, as well as the INMS measurement of \citet{vui07} and the theoretical profiles from \citet{kra09} and \citet{loi14}. Vertical dotted portion of the blue `gradient model' curve shows a plausible alternative profile above 700~km. (b) Spectral region used for our model fits, containing some of the strongest observed \etcn\ lines (black histogram). The best-fitting \etcn\ models from panel (a) are shown with colored curves. (c) Close-up of the region surrounding three of the strongest \etcn\ emission features. \label{fig:models}}
\end{figure}

Contour maps showing the spatial distributions of emission from the observed species are shown in Fig. \ref{fig:maps}, overlaid upon the 221-225~GHz continuum flux (shown in each panel as an orange bitmap). The contour maps were obtained by integrating over the strongest line(s) observed for each species. For \etcn, the strongest six lines in the range 223.9-224.1~GHz (shown in Fig. \ref{fig:models}(b)) were summed to improve the signal-to-noise ratio of the map. For HC$_3$N, only the strongest ($J=26-25$) line was used, and for CH$_3$CN and CH$_3$CCH, integration was performed over the blended multiplets at 239.8~GHz and 222.1~GHz, respectively. The peak signal-to-noise ratios of the resulting maps were as follows: \etcn: 15, HC$_3$N: 40, CH$_3$CN: 17, CH$_3$CCH: 29.

Adopting a similar strategy to \citet{cor14}, the spatially-integrated \etcn\ emission lines in the range 223.9-224.2~GHz were modeled using the line-by-line radiative transfer module of the NEMESIS atmospheric retrieval code \citep{irw08}. There are 13 \etcn\ transitions in this range (see Table \ref{tab:trans}), spanning a broad range of upper-state energies (150-300~K), and providing sufficient information on the \etcn\ abundance and excitation to {adequately} constrain our models. A lack of interloping lines from other species in this range facilitated accurate abundance retrievals. The atmospheric temperature profile was generated from a combination of Cassini CIRS and HASI measurements \citep{fla05,ful05}, and the abundances of nitrogen and methane isotopologues and aerosols were the same as used by \citet{tea13}. Spectral line parameters and partition functions were taken from the JPL catalog. The Lorentzian broadening half-width at 296~K was assumed to be 0.075~cm$^{-1}$\,atm$^{-1}$, with a temperature-dependence exponent of 0.50.

Under the assumption that the model temperature/abundance/aerosol profiles profiles do not vary with latitude and longitude, the integrated flux per spectral channel was calculated using a 10-point trapezium-rule integration from the center of Titan's disk to the edge of the model atmosphere (at an altitude of 1000~km). Initially, least-squares fits to the observed spectra were performed using \etcn\ abundance profiles with constant mixing ratio above a specified cutoff altitude $z_c$ and zero abundance below. In these `step' models, $z_c$ values {in the range $100$-400~km were tried, and the best-fitting abundances for each} are given in Table \ref{tab:models}. Reduced $\chi^2$ values are also given, and the corresponding continuum-subtracted spectral models are shown in Fig. \ref{fig:models}.  

{The $z_c = 300$ and 400~km models (green and magenta solid curves, respectively), provide a near-perfect fit} to the observations, accurately reproducing the intensities, frequencies and widths of all the spectral features attributed to \etcn. The quality of {these fits} confirms that our \etcn\ line assignments are secure. As can be seen from the close-up spectral region in Fig. \ref{fig:models}c, our models show that pressure-broadening wings begin to become significant for $z_c \lesssim200$~km, but no {clear} wings are apparent above the noise level in the observed spectra, indicating that the majority of Titan's \etcn\ must be confined to higher altitudes. {Accordingly, the $z_c = 100$~km model (orange dashed curve) provides a poor fit to the observations.}

Due to vertical mixing, model profiles with an abrupt abundance cutoff are unlikely to represent the true \etcn\ profile, so the spectrum was also fitted using a more realistic two-parameter `gradient' model, in which the abundance varies smoothly as a function of altitude. The free parameters in this model were the abundance at a reference altitude $z_r = 292$~km, and the slope, given as a fraction ($f_H$) of the atmospheric pressure scale height. Condensation of \etcn\ was assumed to occur at altitudes $\lesssim100$~km based on the saturated vapor pressure curve of \citet{lid94}. The best-fitting abundance at $z_r$ was found to be $1.3 \pm 0.4$~ppb, with a retrieved fractional scale height of $f_H = 4.6 \pm 1.1$.  Despite {slightly stronger line wings, this model (blue solid curve) also matches the observations extremely well.}

Vertical column densities ($N$) for the four models are given in Table \ref{tab:models}. {The models that fit the data well have values in the range $(1-5)\times10^{14}$~cm$^{-2}$.}

\begin{table}
\centering
\caption{Best-fitting C$_2$H$_5$CN model abundances and vertical column densities \label{tab:models}}
\begin{tabular}{lcccc}
\hline\hline
Model&Abundance (ppb)&$z_r$$^a$ (km)&$\chi^2$&$N$ (cm$^{-2}$)\\
\hline
Step (100~km)&0.79&100&1.43&$1.3\times10^{15}$\\
Step (200~km)&3.24&200&1.01&$4.6\times10^{14}$\\
Step (300~km)&9.25&300&0.97&$1.7\times10^{14}$\\
Step (400~km)&73.1&400&0.98&$2.1\times10^{14}$\\
Gradient&1.30&292&0.97&$3.6\times10^{14}$\\
\hline
\end{tabular}
\\
\parbox{\columnwidth}{\footnotesize 
\vspace*{1mm}
$^a$Reference altitude for abundance.\\
}
\end{table}

\section{Discussion}
\subsection{\etcn\ Vertical Mixing Ratio Profile}

Confirmation of the presence of ethyl cyanide in Titan's atmosphere provides an important test and validation of the Cassini INMS result. Although a near-perfect fit to our observed ALMA spectrum was obtained using a step model for the \etcn\ VMR profile (with cutoff altitude at 300~km), a smoothly-varying profile is expected to be more realistic. The shape of the retrieved `gradient' profile (shown by the solid blue curve in Fig. \ref{fig:models}a) is constrained primarily by the requirement of a low abundance at $z\lesssim200$~km, and is largely unconstrained at altitudes above about 600~km, where the atmospheric density is low. Similar to the case of HNC \citep{mor12,cor14}, the steep vertical abundance gradient for \etcn\ and its concentration at altitudes $\gtrsim200$~km are consistent with production in the upper atmosphere, combined with a relatively short chemical lifetime that prevents it from building up at lower altitudes through vertical mixing.

A plausible {alternative to the best-fitting gradient profile above 700~km} is indicated with a dotted vertical line in Fig. \ref{fig:models}a, passing through {the $5\times10^{-7}$ abundance at $z=1,100$~km, measured by \citet{vui07}.} The modeled \etcn\ column density above 700~km is small, so this alternative curve also provides an excellent fit to the observed spectrum. The altitude of peak sensitivity of our measurements depends on the assumed VMR profile shape, but for reasonable profiles that do not exceed the abundance measurement of \citet{vui07} at high altitude, the resulting spectra are most sensitive to \etcn\ at altitudes less than 600~km.

In Fig. \ref{fig:models}a, we plot for comparison the \etcn\ VMR profiles predicted by the chemical models of \citet{kra09} and \citet{loi14}. In the $z\approx200$-600~km region from which the majority of our observed emission originates, neither chemical model matches well with our retrieved VMR profiles. The \citet{kra09} model produces \etcn\ through the reaction of excited nitrogen atoms with C$_3$H$_6$, and the resulting stratospheric/mesospheric abundance is up to {three} orders of magnitude lower than implied by our observations. This indicates that a significant \etcn\ production channel is missing from the model. Conversely, \etcn\ abundances calculated by \citet{loi14} exceed those of our retrieved profiles by 1-3 orders of magnitude at altitudes 100-300~km, where our model is quite well constrained. In their model, \etcn\ is produced from the ternary association reaction of CH$_2$CN and CH$_3$ (+ M), which has not been measured in the laboratory. The apparent shortcomings of these models highlight a lack of knowledge regarding {Titan's \etcn\ chemistry. Improved chemical models are currently in development \citep[\eg][]{vui14}.}

The production of \etcn\ in the mid and upper atmosphere on Titan could plausibly be the result of gas-phase chemistry involving binary ion-molecule or neutral-neutral reactions. Production and subsequent release from the surfaces of solid (haze) particles or ices also cannot be ruled out --- \etcn\ is abundant in the environments surrounding young stellar objects, where its formation is generally believed to result from reactions on the surfaces of dust grains \citep[\eg][]{bis07,bot07}. As yet, there are no known neutral gas-phase routes to the formation of ethyl cyanide in low pressure astrophysical environments, and our results identify a clear need for further laboratory and theoretical studies in this regard. Recombination of C$_2$H$_5$CNH$^+$ is known to produce \etcn\ \citep{vig10}, so ion-molecule synthesis of this nitrile ion should also be examined.

\subsection{Molecular Emission Maps}

Combining the derived vertical profile with the observed emission maps, the three-dimensional distribution of \etcn\ relative to other atmospheric species provides important clues regarding its origin.  {The observed molecular contour maps in Fig. \ref{fig:maps} show evidence for significant north-south extension of each species.} This extension is consistent with the presence of emission from near both poles, which is highlighted in the case of HC$_3$N, for which two closely-separated emission components (one over each pole) could plausibly give rise to the observed contour structure. Indeed, \citet{cor14} observed the HC$_3$N $J=40-39$ transition {towards} Titan in November 2013 at slightly higher angular resolution, and identified two clear emission peaks for this molecule, one centered near each pole. Such polar enhancement(s) are characteristic of the abundance patterns seen by Voyager and Cassini in photochemical product species, theorized to result from the combined effects of Titan's atmospheric chemistry and circulation \citep{cou95,tea08,cou10,vin10,tea12}.

The \etcn\ emission map (Fig. \ref{fig:maps}a) shows a surprising spatial distribution, with a significant ($7\sigma$) flux enhancement over the southern pole relative to the north. This is the opposite trend to that observed for the other three species, which show clear enhancements in the northern hemisphere. Differences in stratospheric temperature between Titan's northern and southern hemispheres have been found to be generally $\lesssim15$\% \citep{cou95,fla05,ach08}. Our observed emission lines are relatively invariant to such small temperature changes, so spatial variations in the gas temperature are an unlikely explanation for the observed asymmetries. Molecular abundance peaks near Titan's north pole have previously been identified for HC$_3$N and CH$_3$CCH (during northern winter), so it seems plausible that the emission structures observed with ALMA are due to intrinsic latitudinal abundance variations. As a caveat, the temperatures in Titan's polar mesosphere are known to vary on rapid timescales \citep[\eg][]{tea12,dek14}, and temperature information for the epoch of our observations is lacking, so further work will be needed to firmly establish the relative distributions of these species.

An explanation for the southerly \etcn\ peak is due to the transition of Titan's seasons from northern winter in 2002 to {\bf(late)} northern spring in 2012. During this time, the reversal of Titan's main atmospheric circulation cell is expected to channel fresh photochemical products from mid latitudes towards the south pole \citep{tea12}. Meanwhile, gases transported north during the previous season remain concentrated around the north pole, undergoing slow photochemical destruction. Following the northern winter, species with longer chemical lifetimes should remain in the north for longer while those with shorter lifetimes disappear, and re-appear in the south. {Our maps therefore suggest a shorter chemical lifetime for \etcn\ than for HC$_3$N, CH$_3$CN and CH$_3$CCH. This hypothesis} can be tested by observing the evolution of the distributions of these species over the coming years, particularly during the transition from northern spring to the summer solstice in 2017.

\section{Summary}

Nineteen separate emission features from 27 rotational transitions of \etcn\ were detected in Titan's atmosphere using ALMA archival spectra recorded in July 2012.

Radiative transfer modeling indicates that most of the observed \etcn\ is concentrated at altitudes $>200$~km. This is consistent with production in the upper atmosphere, combined with a relatively short chemical lifetime {that inhibits downward mixing}.  The observed \etcn\ emission maps suggest a higher abundance in the south polar region than the north {(in contrast to HC$_3$N, CH$_3$CN, and CH$_3$CCH, which peak in the north)}, again consistent with a {relatively} short chemical lifetime for \etcn\, and south polar subsidence. {This possibility can be verified through future measurements of temporal variations in the spatial distributions of the observed species with respect to Titan's changing seasons.}

In order to help ascertain the origin of ethyl cyanide on Titan (and in other astrophysical environments), new laboratory studies of possible gas-phase routes to \etcn\ are required.

\acknowledgments
This research was supported by NASA's Planetary Atmospheres and Planetary Astronomy programs, The Goddard Center for Astrobiology, The Leverhulme Trust and the UK Science and Technology Facilities Council. It makes use of ALMA data set ADS/JAO.ALMA\#2011.0.00319.S. ALMA is a partnership of ESO (representing its member states), NSF (USA) and NINS (Japan), together with NRC (Canada) and NSC and ASIAA (Taiwan), in cooperation with the Republic of Chile. The Joint ALMA Observatory is operated by ESO, AUI/NRAO and NAOJ. The National Radio Astronomy Observatory is a facility of the National Science Foundation operated under cooperative agreement by Associated Universities, Inc.

\end{document}